\documentclass[cameraready]{Interspeech}

\title{Rethinking Discrete Speech Representation Tokens for Accent Generation}

\author[affiliation={1}, orcid=0009-0005-1058-8711, equalcontribution]{Jinzuomu}{Zhong}
\author[affiliation={1}, orcid=0000-0000-0000-1111, equalcontribution]{Yi}{Wang}
\author[affiliation={1}, orcid=0000-0000-0000-1111]{Korin}{Richmond}
\author[affiliation={1}, orcid=0000-0000-0000-1111]{Peter}{Bell}

\address{
    $^1$ Centre for Speech Technology Research, University of Edinburgh, UK   
}

\email{jinzuomu.zhong@ed.ac.uk, wang.yi@ed.ac.uk, korin.richmond@ed.ac.uk, peter.bell@ed.ac.uk}

\keywords{accent, discrete speech representation, voice conversion, accent conversion}

\definecolor{darkgreen}{RGB}{0,100,0}
\newcommand{\darkgreen}[1]{\textcolor{darkgreen}{#1}}

\usepackage{comment}

\usepackage{booktabs,tabularx,makecell}
\usepackage{multirow,tipa}
\usepackage{xcolor}
\usepackage{subcaption}
\usepackage[normalem]{ulem}

\begin{document}

\maketitle

\begin{abstract}
    Discrete Speech Representation Tokens (DSRTs) have become a foundational component in speech generation. While prior work has extensively studied phonetic and speaker information in DSRTs, how accent information is encoded in DSRTs remains largely unexplored. In this paper, we present the first systematic investigation of accent information in DSRTs. We propose a unified evaluation framework that measures both \textit{accessibility} of accent information via a novel Accent ABX task and \textit{recoverability} via cross-accent Voice Conversion (VC) resynthesis. Using this framework, we analyse DSRTs derived from several widely used speech representations. Our results reveal that: (1) choice of layers has the most significant impact on retaining accent information, (2) accent information is substantially reduced by ASR supervision; (3) naive codebook size reduction cannot effectively disentangle accent from phonetic and speaker information. 
\end{abstract}

\section{Introduction}

Inspired by the tokenisation schemes employed by text-based Large Language Models (LLMs), discrete speech tokenisation has emerged as an active research area for bridging speech and LLMs. By compressing speech signals or learned continuous representations into a finite set of discrete units, \textit{discrete speech tokens} enable efficient storage of complex acoustic information while remaining inherently compatible with LLM architectures. As a result, they are rapidly becoming established as a foundational representation across a wide range of speech tasks \cite{guo2025recent}.

Directly compressing raw speech signals, however, tends to produce a representation that preserves abundant low-level acoustic detail, making phonetic content difficult to access. Such an approach works well for waveform reconstruction, but places a heavier modelling burden on downstream speech generation tasks \cite{wang2025maskgct}. To address this, \textit{discrete speech representation tokens (DSRTs)}, which are quantised from speech representations, are often used. DSRTs have seen huge success in various speech generation tasks, including Speech Language Models (SpeechLMs) \cite{cui2025recent}, Zero-Shot Text-to-Speech (ZS-TTS) \cite{kharitonov2023speak}, Speech-to-Speech Translation (S2ST) \cite{lee2022direct} and full-duplex dialogue systems \cite{nguyen2023dgslm}.

Despite the significant progress brought by DSRTs, the representation of a speaker's \textit{accent} is largely overlooked in the design, evaluation, and application of the tokens. Prior user perception studies have demonstrated a similarity--attraction effect, whereby listeners prefer accents similar to their own in spoken interaction \cite{dahlback2007similarity}. Unfortunately, ZS-TTS systems are shown to hallucinate accents that differ from those of the reference speakers \cite{zhong2025accentbox}, while spoken dialogue systems still lack broad adoption of benchmarks or evaluation settings that consider appropriate, accent-controlled speech generation \cite{cheng2025voxdialogue}. 

Although some recent ZS-TTS systems have demonstrated some ability to mimic or control the accent of generated speech using DSRTs, how much accent information is encoded in these tokens remains unexplored \cite{du2024cosyvoice, zhang2025vevo, wang2025maskgct}. Existing claims -- such as that naive codebook size adjustment \cite{zhang2025vevo} or ASR supervision \cite{du2024cosyvoice} can facilitate accent control and generation -- lack systematic investigation. Crucially, the accent information actually preserved in these DSRTs remains unquantified, leaving it unclear whether the observed accent generation capabilities in ZS-TTS stem from the representations themselves or are merely a byproduct of large-scale pretraining. To this end, we ask the following research questions: (1) \textit{How do different design choices in DSRTs influence the amount of accent information they encode?} (2) \textit{How can these insights be leveraged to enable more controllable accent generation, e.g. in voice conversion (VC)?}

To answer the research questions, we propose a framework for evaluating DSRTs from both the \textit{recoverability} and \textit{accessibility} perspectives, in terms of accent, speaker, and phonetic information encoded. Existing DSRT evaluation frameworks focus on phonetic and speaker information and rarely consider accent \cite{polyak2021speech}. To the best of our knowledge, this is the first work to incorporate accent in evaluating DSRTs, and to apply ABX to the study
of accent. Specifically, we extend the existing investigation of accessibility to accent information, with a novel accent ABX method that evaluates the discriminability of representations for words uttered in different accents. Additionally, we propose to evaluate the recoverability of accent, speaker, and phonetic information by cross-accent VC, resynthesising with DSRTs from source speakers and speaker IDs from target speakers of different accents as input. 

Using this proposed framework, we observe the following key findings: (1) Accent information is most prominent in mid-early layers of HuBERT, different from speaker or phonetic information distribution. (2) Naive adjustment of codebook sizes provides only very limited disentanglement of accent, speaker and content information. (3) Predominant design of DSRTs for speech generation (quantising a later layer in a speech representation model, or using ASR supervision) discards most accent information. These findings challenge existing claims using naive codebook size adjustment \cite{zhang2025vevo} or ASR supervision \cite{du2024cosyvoice} for accent control and generation.

Finally, based on our findings, we propose DSRT design choices that are appropriate for both \textit{accent-preserving} VC -- preserving source speaker accent, and \textit{accent-adaptive} VC -- adapting to target speaker accent. Both objective and subjective evaluation show superior performance to existing approaches.
\ifcameraready
    Link to demo: \url{https://jzmzhong.github.io/Accent-DSRT}
\else
    We will release links to code and demo in the final version.
\fi


\section{Related Work}

\subsection{Discrete Speech Tokens}
\label{ssec:bg_dst}

Discrete speech tokens can be largely categorised into two groups: \sout{\textit{semantic tokens}}\footnote{A severe misnomer as these tokens contain primarily phonetic information and little semantic information \cite{wells2022phonetic, choi2024self}.} and \textit{acoustic tokens}. We refer to \sout{semantic tokens} as Discrete Speech Representation Tokens (DSRTs) here, naming them strictly by how these tokens are designed and trained, rather than what information some researchers assume them to possess. DSRTs are tokens obtained by quantising learned speech representations, such as HuBERT \cite{hsu2021hubert} and w2v-BERT \cite{chung2021w2vbert}; while acoustic tokens are obtained by directly quantising waveforms. Common quantisation techniques include k-means, Vector Quantisation (VQ) \cite{vandenoord2017neural}, Residual Vector Quantisation (RVQ) \cite{zeghidour2021soundstream}, and Finite-State Quantisation (FSQ) \cite{mentzer2024finite}, with k-means, VQ, and FSQ primarily used for DSRTs, VQ and RVQ primarily used for acoustic tokens.

\subsection{Investigation of Speech Representations}
\label{ssec:bg_ssl}

In terms of of continuous speech representations, most previous work has focused on accessibility, finding that phonetic information peaks in intermediate layers, while acoustic information dominates earlier layers \cite{choi2024self, pasad2021layer}. After finetuning on the ASR task, it has been observed deeper layers become dominated by task-specific information, exhibiting higher phonetic discrimination.

In the investigation of DSRTs, Wells et al.\ \cite{wells2022phonetic} find that DSRTs obtained by k-means on \texttt{hubert-base-ls960} correspond to sub-phonetic events. Yeh et al.\ \cite{yeh2024estimating}, using their information completeness and accessibility framework, found that DSRTs obtained from \texttt{hubert-base-ls960} layer 4 contain more accessible pitch and speaker information, but less accessible phone information, than layer 9. Phonetic and speaker information in both layers consistently decrease with lower bitrates of RVQ codes, refuting claims that content and speaker disentanglement can be achieved by VQ.

\subsection{DSRTs in ZS-TTS and SpeechLMs}
\label{ssec:bg_tts}

In ZS-TTS, researchers found that DSRTs provide more accessible phonetic information than acoustic tokens do, while suffering information loss such as timbre and acoustic environment \cite{wang2025maskgct}. Motivated by such findings, many state-of-the-art (SOTA) models utilise a hierarchical approach, first predicting DSRTs from text or source speech and then predicting acoustic tokens from DSRTs \cite{du2024cosyvoice, lee2025hierspeechpp, wang2025maskgct, zhang2025vevo}. In SpeechLMs, researchers seek to combine DSRTs with acoustic tokens for both accessible and complete/recoverable information, such as SpeechTokenizer \cite{zhang2024speechtokenizer} and Mimi tokeniser \cite{defossez2024moshi}.

When designing such DSRTs, ZS-TTS researchers have introduced some claims that are either unverified or contradicted by speech representation investigations. Vevo \cite{zhang2025vevo} chooses layer 18 of \texttt{hubert-large-ll60k} and claims that reducing the codebook size in VQ leads to natural disentanglement of speaker, style (accent and emotion by their definition) and content, with a codebook size of 32 containing only content, and a codebook size of 8,192 containing only content and style. This claim not only contradicts the findings in \cite{yeh2024estimating} that VQ or k-means with 1024 codebook size still retain significant speaker information, but also lacks verification in terms of accent and emotion. CosyVoice \cite{du2024cosyvoice} proposes to use ``supervised semantic token'', obtained by injecting FSQ in an internal ASR model encoder, for content consistency in the generated speech; how ASR pretraining affects accent information in discrete tokens remains unclear.

\subsection{Accent Control and Generation}
\label{ssec:accent}

Unfortunately, ZS-TTS speech generation systems suffer \textit{accent hallucination}, whereby they generate a hallucinated synthetic accent that deviates from the reference or prompt speech \cite{zhong2025accentbox}. Recent systems that have been specially built for accent generation have not yet utilised DSRTs along with powerful LLM architectures \cite{zhong2025accentbox, xinyuan2025scalable}. General ZS-TTS systems are reported to have certain accent control capabilities, but there has been little or no investigation about accent information in DSRTs. Understanding how accent information is encoded could help build more inclusive ZS-TTS systems for diverse accents.

\subsection{ABX Evaluation}
\label{subsection: ABX Evaluation}

The ABX error rate is a distance-based, model-agnostic evaluation metric, assessing whether the representations pull data of the same category (e.g. sample $x =$ \textipa{[bIt]}, another sample $a =$ \textipa{[bIt]}, of the same triphone category) closer and push dissimilar data (e.g. $x =$ \textipa{[bIt]}, $b =$ \textipa{[bi:t]}) farther in representational space.

Previous work introduced Minimal Pair ABX (MP-ABX) \cite{schatz2013evaluating}, where instances differ only in a central phonetic unit while sharing the same phonetic context. Prior studies \cite{wang2025maskgct, borsos2023audiolm, zhang2025vevo} have used this metric to assess the phonetic discriminability of DSRT tokens.
In addition, ABX has been extended to probe speaker-related information by defining triplets that contrast speaker identity under controlled linguistic content \cite{de2022language}. Nevertheless, to the best of our knowledge, no prior work has systematically employed machine ABX to reveal \textbf{accent} discriminability in speech representations or DSRTs.

\section{Method}
\label{sec:method}

\begin{figure*}[t]
    \centering
    \includegraphics[width=0.8\linewidth]{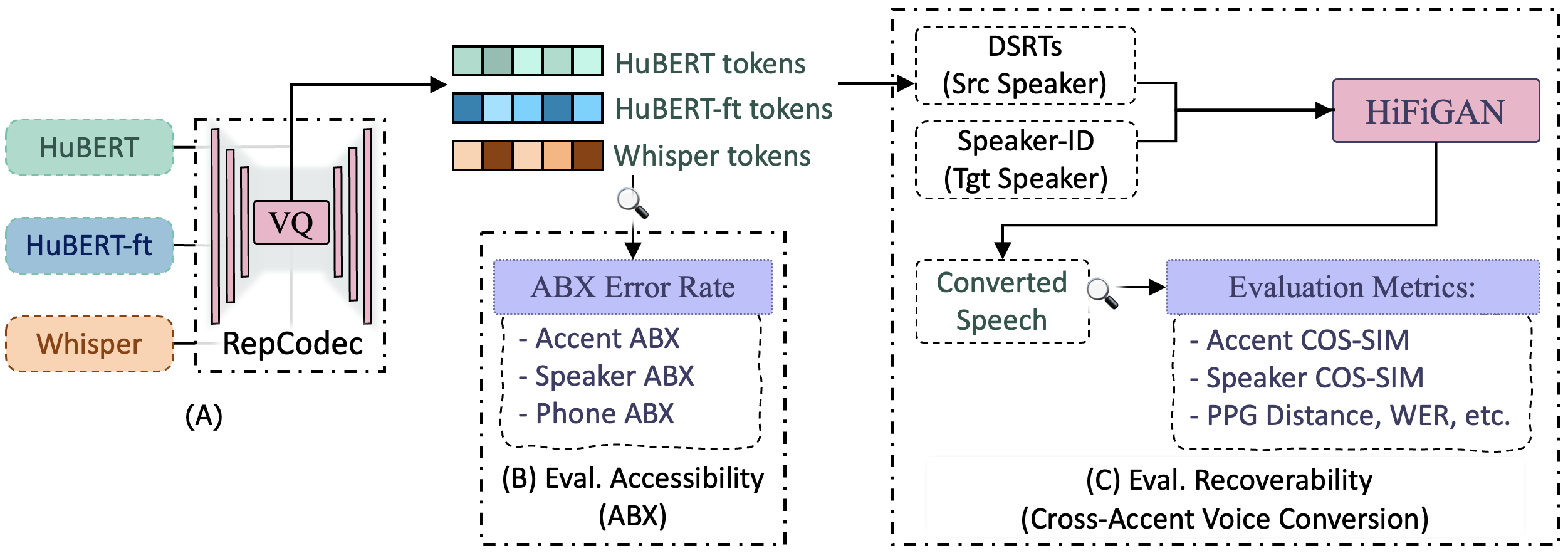}
    \captionsetup{justification=centering}
    \caption{Proposed pipeline for evaluating the \textbf{recoverability} and \textbf{accessibility} of \textit{accent}, \textit{speaker}, and \textit{phonetic} information in various Discrete Speech Representation Tokens (DSRTs).}
    \label{fig:pipeline}
    \vspace{-1em}
\end{figure*}

Inspired by the perspective of separately measuring information completeness and accessibility \cite{yeh2024estimating}, we focus on evaluating DSRTs from both aspects, but in the context of accent generation. Since completeness cannot be measured by resynthesis, we introduce \textit{recoverability} as a task-grounded measure of how much accent, speaker, and phonetic information encoded in DSRTs can be recovered in resynthesised speech. Together with \textit{accessibility}, we propose a pipeline that evaluates DSRTs from both synthesis-facing and representation-facing perspectives.

Figure \ref{fig:pipeline} provides an overview of the proposed pipeline. First, we obtain DSRTs from several commonly used speech representations using RepCodec \cite{huang2024repcodec} with Vector Quantisation (VQ) \cite{vandenoord2017neural} as the quantiser (see Section \ref{ssec:vq-dsrt}). For each DSRT configuration, we then train a unit-to-speech resynthesis model using HiFiGAN \cite{polyak2021speech}. To assess information \textit{recoverability}, we infer unit-to-speech models with DSRTs from a source speaker and speaker ID from a target speaker with a different accent, thereby conducting \textit{cross-accent VC}. The generated speech is evaluated using a combination of objective metrics and subjective listening tests (see Section \ref{ssec:cross-accent-vc}). Finally, to assess information \textit{accessibility}, we directly probe the DSRTs using a range of \textit{ABX} setups (see Section \ref{ssec:abx}).

\subsection{Discrete Speech Representation Tokens}
\label{ssec:vq-dsrt}

We choose three speech representation models to obtain Discrete Speech Representation Tokens (DSRTs): HuBERT; HuBERT finetuned for ASR (HuBERT-ft); and Whisper. We include HuBERT \cite{hsu2021hubert} as it provides the most widely used speech representations in various speech tasks such as SpeechLM and TTS \cite{guo2025recent, cui2025recent}. Recently, there has been a growing trend toward using ASR-based speech representations for obtaining DSRTs. Therefore, we include HuBERT-ft and Whisper \cite{radford2023robust} as two representative ASR-based representations, with Encoder-only and Encoder-Decoder architectures, respectively. All three speech representations investigated, despite being at least three years old, are still widely used in current speech LLMs, evidenced by their usage in speech tokenisers of 19 systems \cite[Table 1]{cui2025recent}.

Following RepCodec \cite{huang2024repcodec}, we use VQ-VAE \cite{vandenoord2017neural} to discretise the speech representations. The model consists of three modules: Encoder and Decoder, which are both convolution layers with residual paths, and a Vector Quantisation module that quantises latent representations from Encoder output into a series of discrete tokens. The model is trained to reconstruct the speech representation, with additional Exponential Moving Average (EMA) optimisation to gradually update the codebook. For model details, we refer readers to the original paper \cite{huang2024repcodec}. 

\subsection{Cross-Accent Voice Conversion}
\label{ssec:cross-accent-vc}

After extracting DSRTs, we train a unit-to-speech HiFiGAN model for each DSRT configuration, following \cite{polyak2021speech}. In prior work, information recoverability is typically evaluated using resynthesis or Voice Conversion (VC) on General American English datasets, with evaluation focusing primarily on intelligibility and speaker similarity, while accent is largely ignored.

We train unit-to-speech models on data covering multiple accents to assess the generalisability of DSRTs across accents. We then perform cross-accent VC in inference, by conditioning the model on DSRTs from a source speaker and a target speaker ID with a different accent. Afterwards, during evaluation, following \cite{zhong2025pairwise}, we explicitly assess accent, speaker, and phonetic similarity in the converted speech, against ground truth reference clips. Utterance-level embeddings from Accent Identification (AID) and Speaker Verification (SV) models are used to calculate cosine similarity, which serves as a proxy for perceived accent and speaker similarity. Phonetic Posteriorgrams (PPGs) are extracted, aligned, and then used to calculate phonetic distances as a proxy for perceived phonetic similarity, and to some degree, intelligibility.

By analysing the similarity of the generated speech to the source and target speech along these dimensions, we can determine whether accent information is primarily derived from the source DSRTs or overridden by the target speaker identity, thereby quantifying the amount of accent information preserved in the DSRTs.

\subsection{Accent, Speaker, and Phonetic ABX}
\label{ssec:abx}

We choose ABX as a model-free method to estimate the accessibility of accent information, as well as phone and speaker information in DSRTs. Table \ref{tab:abx_conditions} lists the three different ABX conditions we use to evaluate accent, speaker, and phonetic information accessibility.

\begin{table}[h!]
\centering
\captionsetup{justification=centering}
\caption{Different ABX conditions for selecting $(a,b,x)$ triplets to evaluate accent, speaker, and phonetic information accessibility. $^*$: \textit{context} refers to the previous and next phones.}
\label{tab:abx_conditions}
\vspace{-0.5em}
\begin{tabularx}{\columnwidth}{
  l c c c
}
\toprule
\makecell[c]{ABX\\type} & \makecell[c]{on\\$(A = X \neq B)$}
 & \makecell[c]{by\\$(A = B = X)$}
 & \makecell[c]{across\\$(A \neq X)$} \\
\midrule
\makecell{Accent}  & accent  & word            & speaker \\ \midrule 
\makecell{Speaker} & speaker & word, accent   & --- \\ \midrule 
\makecell{Phone}   & phone   & context$^*$  & speaker \\
\bottomrule
\end{tabularx}
\vspace{-1em}
\end{table}

When constructing $(a,b,x)$ triplets, a condition can be described as ABX \textbf{ON} the category of interest, \textbf{BY} the category of controlled constant among triplets, \textbf{ACROSS} the category of free variables among triplets. The classic Phone ABX select triplets ON phone BY context, where 1) $x$ and $a$ share the same phone, with $b$ of a different phone, and 2) $x$, $a$, and $b$ share identical phonetic context, e.g. bit \textipa{[bIt]} vs beat \textipa{[bi:t]}. To evaluate how prominent phonetic information is, compared to other variables, such as speaker, most Phone ABX would also be conditioned ACROSS speaker, with $a$, $b$ coming from the same speaker, and $x$ from a different speaker, e.g. $x =$ \textipa{[bIt]} by speaker$_1$, $a =$ \textipa{[bIt]} by speaker$_2$, $b =$ \textipa{[bi:t]} by speaker$_2$. 

In the proposed Accent ABX, instances $a$ and $x$ share the same accent, while instance $b$ differs in accent. Triplets $(a,b,x)$  are constructed to share identical lexical content rather than phonetic context, as accent distinctions arise from accent-dependent realisations of the same words or lexical units, which may involve different phone sequences. We further require that $a$, $b$, and $x$ come from different speakers to avoid trivial similarity between $a$ and $x$ due to shared speaker identity, e.g. a triplet $(a,b,x)$ comprosing $x=$ ``water'' spoken by a Scottish speaker$_1$, $a=$ ``water'' by a Scottish speaker$_2$, and $b=$ ``water'' by a Southern English speaker$_3$.


In practice, words uttered by different speakers differ due to both accent-induced pronunciation variation and speaker-specific pronunciation patterns. While standard ABX aggregates scores over all valid triplets, we adopt a word selection scheme to improve the sensitivity of Accent ABX. We first select the 100 most frequent words in the training data, compute ABX scores for each (accent $A$, accent $B$, word) combination, and retain the 10\% with the lowest scores as the most accent-discriminative. The choice of most frequent words is to ensure there are enough triplets to sample from in Accent ABX, while the selection of most accent-discriminative combinations is to avoid conducting Accent ABX on words that are pronounced similarly in two distinct accents. The combination selection is performed using continuous features from GenAID \cite{zhong2025accentbox}, a supervised Accent Identification (AID) model finetuned from XLSR \cite{babu2021xls}, to ensure effective word selection and a fair comparison among DSRTs.

\begin{table}[t!]
\centering
\captionsetup{justification=centering}
\caption{Samples of most accent-discriminative (Accent $A$, Accent $B$, Word) combinations selected, for Accent ABX.}
\label{tab:accent_triplet}
\vspace{-0.5em}
\begin{tabular}{lll}
\toprule
Accent A & Word & Accent B \\
\midrule
SouthernEnglish & first & Scottish \\
NorthernEnglish & first & Scottish \\
SouthernEnglish & Scottish & Irish \\
SouthernEnglish & however & Scottish \\
SouthernEnglish & work & Scottish \\
NorthernEnglish & however & Scottish \\
Irish & last & Scottish \\
Scottish & Scottish & Irish \\
NorthernEnglish & Scottish & Irish \\
SouthernEnglish & work & Irish \\
NorthernEnglish & their & Scottish \\
SouthernEnglish & first & Irish \\
NorthernEnglish & however & Irish \\
SouthernEnglish & however & Irish \\
SouthernEnglish & their & Scottish \\
SouthernEnglish & never & Scottish \\
SouthernEnglish & year & Scottish \\
SouthernEnglish & next & Irish \\
SouthernEnglish & Mr. & Scottish \\
NorthernEnglish & year & Scottish \\
NorthernEnglish & other & Irish \\
SouthernEnglish & other & Irish \\
Scottish & first & SouthernEnglish \\
SouthernEnglish & from & Irish \\
Scottish & first & NorthernEnglish \\
\bottomrule
\end{tabular}
\vspace{-1em}
\end{table}



Table \ref{tab:accent_triplet} shows the first 25 selected combinations as an example. The table reveals a few well-established phonetic dimensions that underlie accent differences, successfully captured by our data-driven selection scheme.
First, rhoticity emerges as a dominant cue. Words such as \emph{first} and \emph{work} sharply distinguish Southern or Northern English from Scottish and Irish accents, reflecting the systematic presence of post-vocalic [r] in Scottish and Irish English and its absence in Southern English. The table shows that our metric takes good advantage of using rhoticity difference.
Second, differences in vowel quality, particularly in stressed lexical vowels, play a 
central role. For instance, \emph{first} and \emph{year} involve the near-front 
vowel spaces, which are realised as long centralised vowels in Southern English but shift 
toward more open or fronted realisations in Scottish and Irish English. 
Third, consonantal realisation, especially of [t], contributes substantially to 
discrimination. Words such as \emph{scottish} and \emph{last} contain environments where 
Scottish English strongly favours glottal or lenited realisations of [t], in contrast to 
the more alveolar realisations found in Southern and Irish English.
Finally, patterns of vowel reduction and weak-form realisation further differentiate 
accents. Function words such as \emph{however}, \emph{other}, and \emph{from} show strong 
schwa reduction in Southern English, whereas Irish and Scottish English tend to preserve 
fuller vowel qualities. These differences in reduction strategies, particularly in 
unstressed positions, amplify accent contrasts and explain why relevant accent-word combinations are picked in the table.

\section{Experiments}
\label{sec:exp}

\subsection{Obtaining DSRTs}
\label{ssec:exp_dsrt}

We extract continuous speech representations from \textit{various layers} of \texttt{HuBERT}\footnote{\url{https://huggingface.co/facebook/hubert-large-ll60k}}, \texttt{HuBERT-ft}\footnote{\url{https://huggingface.co/facebook/hubert-large-ls960-ft}} and \texttt{Whisper}\footnote{\url{https://huggingface.co/openai/whisper-medium.en}}. All three speech representation models share the same Encoder architecture (24 Transformer layers) and are trained on English-only data. Here, we focus on investigating how layer choice and ASR pretraining affect the information encoded in speech representations, and leave the investigation of how multilingual pretraining or larger-scale pretraining data/model to future work.

The extracted representations are then discretised using RepCodec\footnote{\url{https://github.com/mct10/RepCodec}}, which we train on the \texttt{train-clean-100} subset of LibriSpeech\footnote{\url{https://www.openslr.org/12}}, similar to \cite{zhang2025vevo, huang2024repcodec}, with \textit{various codebook sizes}. This subset contains 100 hours of speech with "accents closer to US English" \cite{panayotov2015librispeech}. RepCodec is trained for 200,000 steps with the same hyperparameters in \cite{huang2024repcodec}.

\subsection{Evaluation Dataset}
\label{ssec:exp_data}

To evaluate information recoverability and accessibility with respect to accent, we use the VCTK corpus\footnote{\url{https://datashare.ed.ac.uk/handle/10283/3443}}, which provides relatively broad coverage of native English accents. Based on the provided country labels and detailed region descriptions, we manually group speakers into 13 accent regions, as some coarse country-level labels (e.g. ``US'', ``England'') conflate multiple distinct regional accent varieties.

We withhold 3--4 speakers from each accent region (50 speakers in total, with balanced gender and accent distribution) for testing. For the remaining speakers, we use 4 accent regions that have sufficient speakers (45 speakers in total) for training unit-to-speech HiFiGAN in Section~\ref{ssec:cross-accent-vc} and for selecting accent-discriminative words in Section~\ref{ssec:abx}. Out of the 13 accent regions, 4 are seen in HiFiGAN training and Accent ABX word selection, 5 are likely seen by RepCodec in LibriSpeech given they are North American accents, and the remaining 4 are unseen at any stage, allowing for analysis on the generalisability of DSRTs across accents.

\subsection{Evaluating Information Recoverability}
\label{ssec:exp_vc}

To evaluate information recoverability, we train unit-to-speech HiFiGAN models\footnote{\url{https://github.com/facebookresearch/speech-resynthesis}} for 100,000 steps with the same hyperparameters as in \cite{polyak2021speech}, using 4 accent regions (45 speakers), as described in Section~\ref{ssec:exp_data}. During inference, we perform cross-accent VC using source (\textit{src}) DSRTs from 12 accent regions (46 speakers, excluding Southern English), together with speaker IDs from 4 Southern English speakers as target (\textit{tgt}) voices. We fix the target voices to one accent region to control for target-side accent variability, enabling a clearer analysis of how accent information is preserved in the source DSRTs.

For \textbf{objective evaluation} of the converted speech, we adopt metrics recommended by \cite{zhong2025pairwise}. (1) \textit{Accent similarity}: We calculate cosine similarity of accent embeddings (\textbf{Accent COS-SIM} or \textbf{A-SIM}) extracted from \texttt{GenAID}\footnote{\url{https://github.com/jzmzhong/GenAID}} \cite{zhong2025accentbox}. \texttt{GenAID} is trained with explicit speaker-accent disentanglement, preserving little speaker information in the extracted accent embeddings. (2) \textit{Speaker similarity}: We calculate cosine similarity of speaker embeddings (\textbf{Speaker COS-SIM} or \textbf{S-SIM}) extracted from \texttt{WavLM}\footnote{\url{https://huggingface.co/microsoft/wavlm-base-plus-sv}} \cite{chen2022wavlm}. These speaker embeddings encode some accent information, evidenced by improved performance in AID when transfer learned from SV models \cite{zuluagagomez2023commonaccent}. (3) \textit{Phonetic similarity}: We calculate Jensen-Shannon distance between aligned PPGs\footnote{\url{https://github.com/interactiveaudiolab/ppgs}} (\textbf{PPG Distance}), extracted from a phone recognition model \cite{churchwell2024ppgs}. It is worth noting that PPG distance, primarily designed for pronunciation distance, is sensitive to accent information, as accent can be characterised partially as different realisations of the same phoneme. (4) \textit{Intelligibility}: We calculate Word Error Rate (\textbf{WER}) using \texttt{whisper-medium.en} transcriptions, compared against ground-truth transcriptions. Despite its evident accent bias \cite{jahan2025unveiling}, WER is included here due to its broad adoption in previous DSRT evaluation frameworks to reflect how DSRTs are commonly selected.

We use the first 24 utterances per speaker, which is an elicitation paragraph shared across speakers, to calculate Accent COS-SIM, Speaker COS-SIM, and PPG distance. In this way, we allow for similarity/distance calculation to either source speech that provides DSRTs or target speech that provides the speaker ID of the same content, removing the influence of content on these metrics. As for WER, we randomly choose 24 utterances per speaker to cover diverse content.

For clearer visualisation, we annotate each metric with a practical lower bound (chance-level performance) and a practical upper bound (best possible performance), which anchor the visual scale and facilitate interpretation of values.
For Accent COS-SIM, PPG, and WER, the lower bound measures the source speaker ground truth with respect to target speaker ground truth, and the upper bound measures the source speaker copy-synthesis with respect to source speaker ground truth.
For Speaker COS-SIM, the lower bound measures similarity between source speaker ground truth and target speaker ground truth, while the upper bound measures similarity between target speaker copy-synthesis and target speaker ground truth.

Finally, we select several DSRTs to evaluate their performance in both accent-preserving VC and accent-adaptive VC, respectively, and conduct \textbf{subjective listening tests} on the converted speech. Three sets of 5-point Similarity MOS (SMOS) listening tests are conducted, with 15 listeners for each listening test, recruited on Prolific\footnote{\url{https://www.prolific.com/}} from the target accent regions.
To evaluate accent similarity between the generated speech and the source speech (\textbf{Accent SMOS (\textit{src})}) in accent-preserving VC, we randomly sample 20 utterances per selected DSRT, with speaker pairs drawn randomly from all 44 possible pairs (mapping from 11 American Midwest, Scottish, and Oceanian speakers to 4 Southern English speakers, covering both seen and unseen source accents). This results in 300 ratings per system per source accent region.
To evaluate accent similarity between the generated speech and the target accent reference clip (\textbf{Accent SMOS (\textit{tgt})}) in accent-adaptive VC, we randomly sample 30 utterances per selected DSRT, again using random speaker pairs from the same 44 possible mappings. This results in 450 ratings per system.
To evaluate speaker similarity between the generated speech and the target speaker reference clip (\textbf{Speaker SMOS (\textit{tgt})}) in both accent-preserving and accent-adaptive VC, we randomly sample 15 utterances per selected DSRT from the same pool of 44 speaker pairs, yielding 225 ratings per system.

\subsection{Evaluating Information Accessibility}
\label{ssec:exp_abx}

Information accessibility is evaluated only on the accents seen during HiFiGAN training, ensuring sufficient numbers of speakers for each accent.  As described in Section~\ref{ssec:abx}, a training set is required to select high-frequency words and accent-discriminative combinations, while a separate test set is used for ABX evaluation to prevent speaker leakage. Accents outside the four seen accents contain no more than four speakers each, severely limiting the number of valid ABX triplets. 

We follow the same training–test partition as in HiFiGAN experiments for simplicity. Accent ABX is conducted using this partition, whereas phone and speaker ABX are computed only on the test split, as they do not require a training phase. \textcolor{black}{Test experiments used all utterances in the split, except for the first 24 utterances to avoid identical sentences across all train and test speakers, which could bias the ABX scores toward lexical-context matching instead of reflecting true accent or speaker discrimination.}
Phone-level alignment (phone boundaries and labels) required by phone ABX, were obtained through the Montreal Forced Aligner with the English ARPABET dictionary \cite{mfa_english_us_arpa_acoustic_2024,gorman2011prosodylab}.

We treat the accent ABX score computed on HuBERT layer 6 continuous features as an upper bound, as this layer exhibits the strongest accent-related information. Similarly, we use the phonetic and speaker ABX scores computed on HuBERT-ft layers 3 and 9 continuous features as upper bounds, respectively, since these layers demonstrate the highest phonetic and speaker recoverability in Section~\ref{sec:results}. In addition, we evaluate the accent ABX error rate of the accent-discriminative GenAID features, which achieves a low error rate of 9.87\%, validating the effectiveness of the proposed metric.

\begin{figure*}[t]
    \centering

    \begin{subfigure}[b]{\linewidth}
        \centering
        \includegraphics[width=0.875\linewidth, trim = 108 370 128 30, clip]{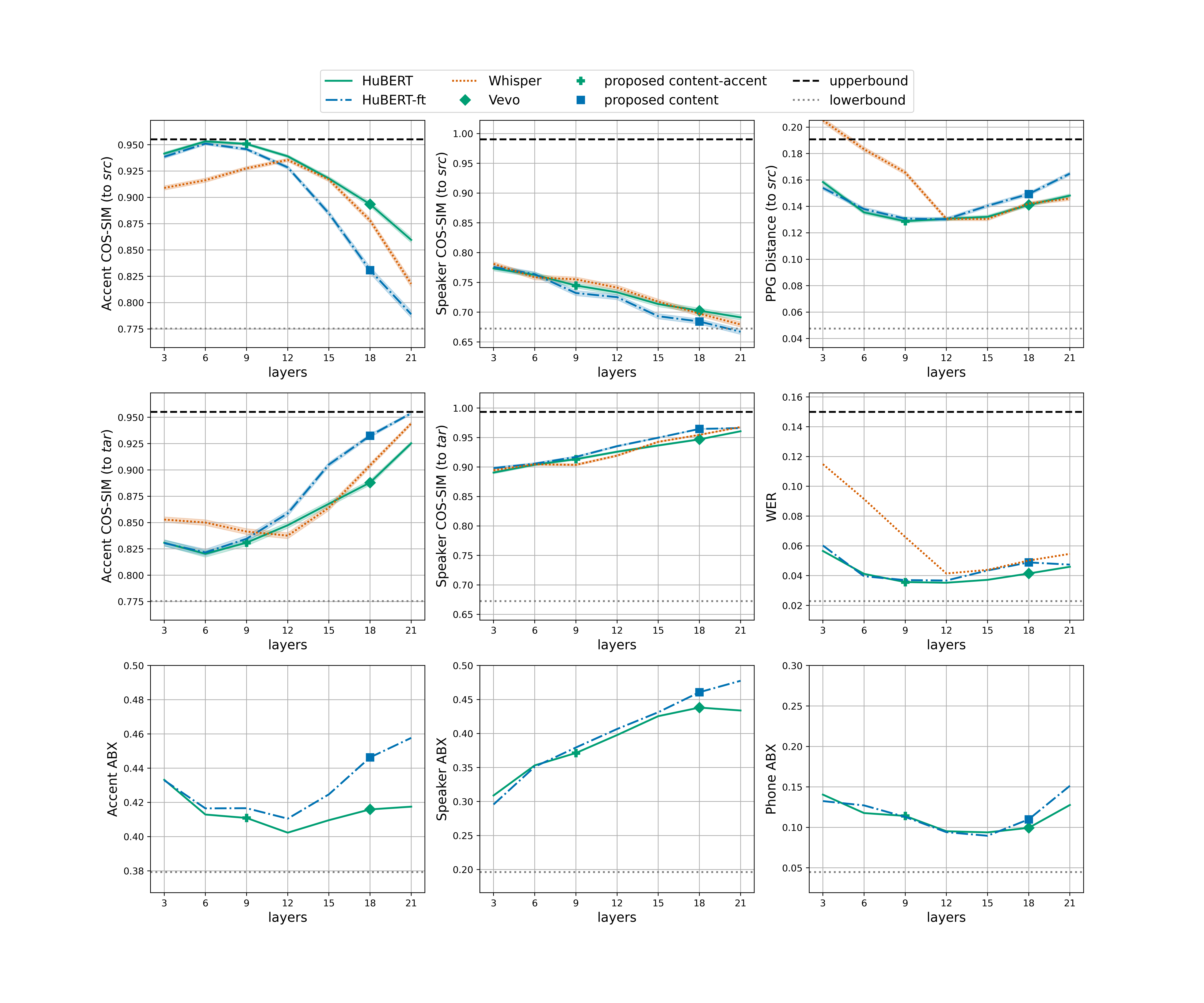}
        \vspace{-0.25em}
        \caption{Cross-accent VC evaluation results for information \textbf{recoverability}.}
        \label{fig:main_plot_subfig1}
    \end{subfigure}
    
    \vspace{0.25em}
    
    \begin{subfigure}[b]{\linewidth}
        \centering
        \includegraphics[width=0.875\linewidth, trim = 108 45 128 375, clip]{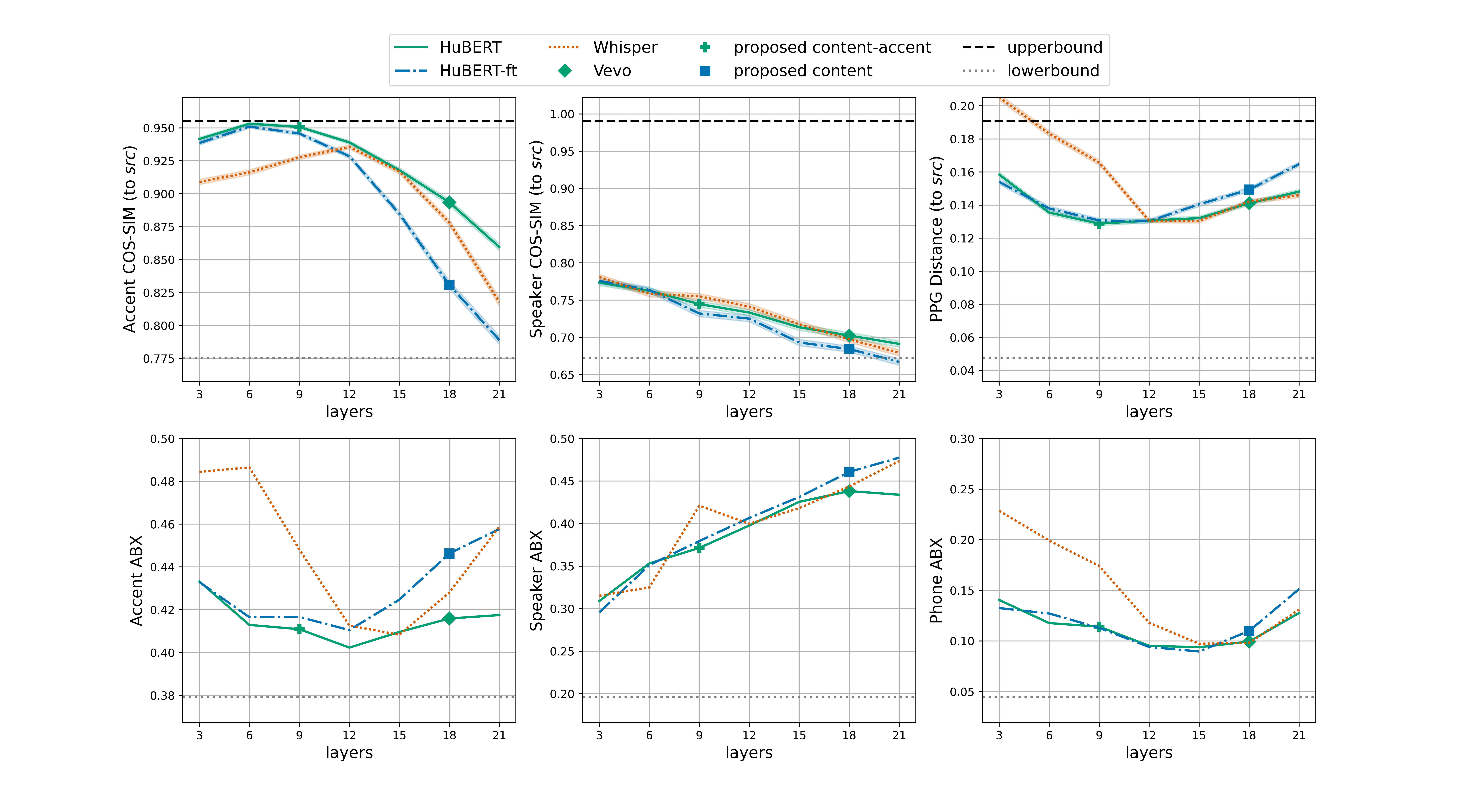}
        \vspace{-0.25em}
        \caption{ABX evaluation results for information \textbf{accessibility}.}
        \label{fig:main_plot_subfig2}
    \end{subfigure}
    
    \vspace{-0.5em}
    \captionsetup{justification=centering}
    \caption{Accent, speaker, and phonetic information in DSRTs obtained across different \textbf{layers} and \textbf{continuous speech representations}.\\ All codebook sizes and code dimensions are set to 1024 for fair comparison.}
    \label{fig:main_plot}
    \vspace{-1em}
\end{figure*}

\begin{figure*}[t]
    \centering
    \begin{subfigure}[b]{\linewidth}
        \centering
        \includegraphics[width=0.875\linewidth, trim = 108 370 128 30, clip]{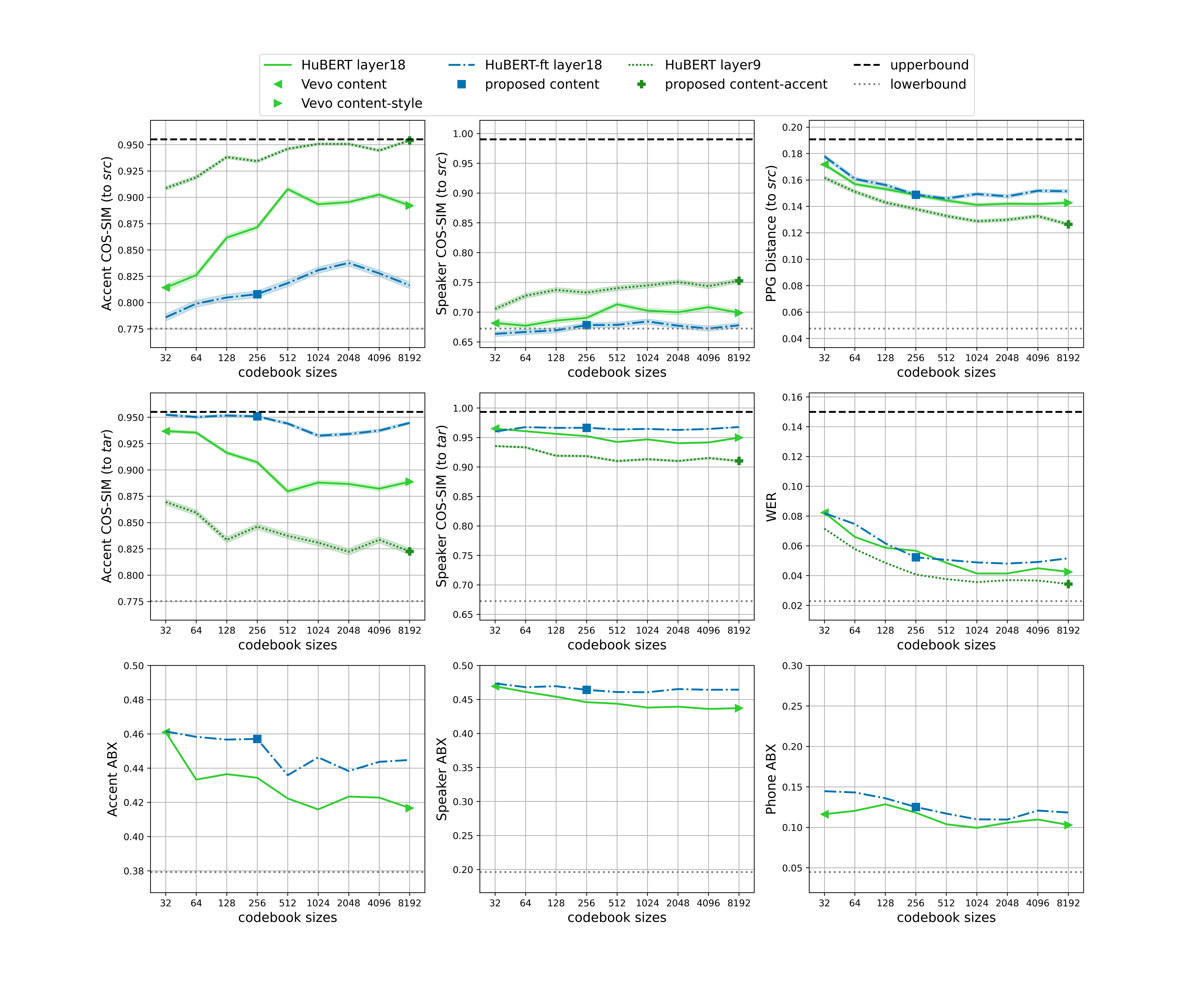}
        \vspace{-0.25em}
        \caption{Cross-accent VC evaluation results for information \textbf{recoverability}.}
        \label{fig:codebooksize_analysis_subfig1}
    \end{subfigure}
    
    \vspace{0.25em}
    
    \begin{subfigure}[b]{\linewidth}
        \centering
        \includegraphics[width=0.875\linewidth, trim = 108 85 128 710, clip]{figs/summary_plots_5_final3_4.png}
        \vspace{-0.25em}
        \caption{ABX evaluation results for information \textbf{accessibility}.}
        \label{fig:codebooksize_analysis_subfig2}
    \end{subfigure}
    
    \vspace{-0.5em}
    \captionsetup{justification=centering}
    \caption{Accent, speaker, and phonetic information in different DSRTs across \textbf{codebook sizes}.}
    \label{fig:codebooksize_analysis}
    \vspace{-0.5em}
\end{figure*}

\section{Results}
\label{sec:results}

\subsection{Layer Choice Matters for Accent}
\label{ssec:choice_of_layers}

Layer choice has a prominent impact on the recoverability and accessibility of information in HuBERT DSRTs, shown in Figure~\ref{fig:main_plot} (\darkgreen{dark green} curve). 
Notably, accent, speaker and phonetic information are distributed differently across layers.
(1) \textit{Accent} information recoverability, reflected by Accent COS-SIM w.r.t.\ source speaker, is most prominent in \textit{mid–early} HuBERT layers (L6 \& L9), and decreases in earlier or latter layers, indicating that accent cues emerge after low-level acoustic processing but are progressively abstracted away in higher layers.
(2) \textit{Speaker} information is most recoverable in \textit{early} HuBERT layers (L3), indicated by Speaker COS-SIM to the source speaker, and decreases monotonically with depth. Overall, speaker information is limited, as cross-accent VC aims to generate voice similar to that of the target speaker with speaker ID conditioning. (3) \textit{Phonetic} information, measured by PPG Distance to the source speech, is most complete in \textit{middle} layers (L9 \& L12). Both earlier and later layers show reduced phonetic completeness, reflecting a trade-off between raw acoustic detail and higher-level abstraction. 
(4) Information \textit{accessibility}, measured by ABX error rates (see Figure~\ref{fig:main_plot_subfig2}), reveals weak accent accessibility for HuBERT across layers, despite clear inter-layer variations. Accent accessibility peaks at L12, which differs from the recoverability peak at L6 or L9. This discrepancy highlights the necessity of resynthesis-based recoverability evaluation when assessing DSRTs for accent generation: while ABX accessibility captures relative trends, it cannot reliably identify optimal layers for synthesis.

\subsection{ASR Supervision Removes Accent Information}
\label{ssec:choice_of_representation}

ASR supervision reduces the recoverability and accessibility of accent information in DSRTs across layers, shown in Figure~\ref{fig:main_plot} (\textcolor{blue}{blue} curve for HuBERT-ft, \textcolor{orange}{orange} curve for Whisper). (1) HuBERT consistently shows higher accent recoverability and accessibility than HuBERT-ft or Whisper, particularly in latter layers, which are more affected by ASR supervision. (2) Whisper exhibits a peak in accent and phonetic information at \textit{intermediate} layers (L12 \& L15), different from the distribution of information in HuBERT or HuBERT-ft.

\begin{table*}[t!]
\centering
\captionsetup{justification=centering}
\caption{Objective evaluation of proposed content and content-accent tokens, compared with content and content-style tokens from Vevo \cite{zhang2025vevo}. Differences in DSRT design choices are marked in \textcolor{red}{red}, with best performances marked in \textbf{bold}.\\ $*$: For content tokens, we want little accent information recovered from DSRTs, i.e. low A-SIM (\textit{src}) and high A-SIM (\textit{tar}). For content-accent or content-style tokens, we want the opposite.}
\label{tab:objective_results}
\vspace{-0.5em}
{

\centering
\begin{tabular}{cccccccccc}
    \toprule
    & Token type & \begin{tabular}[c]{@{}c@{}}Representation\\ \& Layer\end{tabular} & \begin{tabular}[c]{@{}c@{}}Codebook\\ size\end{tabular} & \begin{tabular}[c]{@{}c@{}}A-SIM\\ (\textit{src})$^*$\end{tabular} & \begin{tabular}[c]{@{}c@{}}A-SIM\\ (\textit{tgt})$^*$\end{tabular} & \begin{tabular}[c]{@{}c@{}}S-SIM\\ (\textit{src})$\downarrow$\end{tabular} & \begin{tabular}[c]{@{}c@{}}S-SIM\\ (\textit{tgt})$\uparrow$\end{tabular} & PPG$\downarrow$ & WER$\downarrow$ \\ \midrule
    \multirow{2}{*}{Vevo} 
    & Content & HuBERT L18 & 32 & 0.8143 & 0.9369 & 0.6817 & 0.9651 & 0.1718 & 0.0824 \\
    & Content-style & HuBERT L18 & 8192 & 0.8923 & 0.8888 & \textbf{0.6989} & \textbf{0.9499} & 0.1427 & 0.0425 \\\midrule
    \multirow{2}{*}{Proposed}
    & Content & HuBERT\textcolor{red}{-ft} L18 & \textcolor{red}{256} & \textbf{0.8081} & \textbf{0.9509} & \textbf{0.6782} & \textbf{0.9667} & \textbf{0.1488} & \textbf{0.0523} \\
    & Content-accent & HuBERT \textcolor{red}{L9} & 8192 & \textbf{0.9541} & \textbf{0.8227} & 0.7529 & 0.9104 & \textbf{0.1265} & \textbf{0.0344} \\
    \bottomrule
\end{tabular}
}
\vspace{-1em}
\end{table*}

\begin{table*}[t!]
\centering
\captionsetup{justification=centering}
\caption{Subjective evaluation of proposed content and content-accent tokens, compared with content and content-style tokens from Vevo \cite{zhang2025vevo}, with best performances marked in \textbf{bold}. 95\% confidence intervals are reported.}
\label{tab:subjective_results}
\vspace{-0.5em}
\begin{tabular}{lcccc}
\toprule
\textit{Accent-preserving VC}    & \multicolumn{3}{c}{Accent SMOS (\textit{src})$\uparrow$}                        & \multirow{2}{*}{Speaker SMOS (\textit{tgt})$\uparrow$} \\ \cmidrule{2-4}
                        & American Midwest   & Oceanian           & Scottish           &                                     \\ \midrule
Vevo content-style      & 2.39±0.15          & 2.81±0.15          & 3.16±0.14          & \textbf{3.18±0.17}                  \\
Proposed content-accent & \textbf{3.61±0.14} & \textbf{3.39±0.14} & \textbf{3.59±0.13} & 2.65±0.18                           \\ \midrule \midrule
\textit{Accent-adaptive VC}      & \multicolumn{3}{c}{Accent SMOS (\textit{tgt})$\uparrow$}                        & \multirow{2}{*}{Speaker SMOS (\textit{tgt})$\uparrow$} \\ \cmidrule{2-4}
                        & \multicolumn{3}{c}{Southern English}                         &                                     \\ \midrule
Vevo content            & \multicolumn{3}{c}{2.82±0.12}                                & 3.40±0.17                           \\
Proposed content        & \multicolumn{3}{c}{\textbf{3.15±0.12}}                       & \textbf{3.60±0.16}                  \\ \bottomrule
\end{tabular}
\vspace{-1em}
\end{table*}

\subsection{Reducing Codebook Size Provides Limited Disentanglement}
\label{ssec:reducing_cbz_no_disentangle}

We then vary codebook sizes for three selected layers: HuBERT L18 (a common choice, as in Vevo \cite{zhang2025vevo}), HuBERT-ft L18 (a layer with little accent information while retaining high phonetic information), and HuBERT L9 (a layer with most accent information), shown in Figure~\ref{fig:codebooksize_analysis}.
(1) Focusing on accent recoverability, varying codebook sizes has a markedly smaller effect than changing representation models or layers. For each layer, increasing codebook sizes from 32 to 2048 consistently improves accent and phonetic recoverability across all inspected DSRTs. However, the recoverability gap induced by codebook sizes variation within a layer is substantially smaller than that induced by changing layers at a fixed codebook sizes. 
(2) Speaker and phonetic recoverability also increase with larger codebook sizes. Taking HuBERT DSRTs as an example, Accent COS-SIM drops sharply as codebook sizes decreases from 1024 to 32, while PPG distance also rises 20.4\%. Despite speaker identity being partially suppressed by the injected speaker ID during resynthesis, speaker recoverability also declines. The parallel degradation across accent, speaker, and phonetic metrics suggests minimal disentanglement when reducing codebook size. The VQ bottleneck acts more as a lossy compressor for all information rather than a filter for specific features.

\subsection{Limitations of Commonly Used DSRTs in Speech Generation}
\label{ssec:limitations_commonly_DSRTs}

Based on the above findings, we highlight the limitations of commonly used DSRTs here. (1) Vevo’s choice of using DSTRs based on HuBERT L18 with codebook size 32 as content tokens and codebook size 8,192 as content-style tokens \cite{zhang2025vevo} is suboptimal. Significant accent information is likely already lost and unrecoverable in HuBERT L18 - leading to insufficient accent information encoded in content-style tokens. Since naive reduction of codebook size cannot remove accent without harming phonetic information, content tokens achieve little accent information at the cost of intelligibility in resynthesised speech. (2) ASR-supervised DSRTs, adopted by \cite{du2024cosyvoice}, exhibit lower peak accent and phonetic recoverability than HuBERT DSRTs. Similar to content-style tokens, ASR-supervised DSRTs cannot encode accent sufficiently.

\subsection{Proposed Content and Content-Accent Tokens}
\label{ssec:proposed_content_accent_tokens}

To enable better accent generation and control, we propose \textit{content-accent} and \textit{content} tokens for accent-preserving and accent-adaptive VC, respectively. A comparison between the proposed tokens and the tokens used in Vevo, together with the objective evaluation metrics, is presented in Table~\ref{tab:objective_results}. The listening test results are reported in Table~\ref{tab:subjective_results}, which further validate the objective findings.

Overall, the proposed content and content-accent tokens outperform Vevo's content and content-style tokens across most metrics, particularly in generating the desired accent while maintaining high content consistency and intelligibility. These results suggest improved disentanglement between accent and content. However, while the proposed content tokens achieve high speaker similarity to the target speaker in accent-adaptive VC, the proposed content-accent tokens perform worse in accent-preserving VC. This indicates that accent information remains partially entangled with speaker characteristics in the content-accent tokens.

The lower speaker similarity observed for content-accent tokens may also be influenced by speaker--accent entanglement in both objective and subjective evaluations. As discussed in Section~\ref{ssec:exp_vc}, the speaker embeddings used in the system encode some accent information. Consequently, when the generated speech more closely matches the source speaker's accent, the measured similarity to the target speaker may decrease. A similar phenomenon may occur in the subjective evaluation. Although listeners were instructed to ignore accent when rating speaker similarity, accent differences may still influence perceived speaker identity, as listeners often associate distinct accents with different speakers. This raises the broader question of whether accent and speaker characteristics can be fully disentangled in practice, particularly given that training data rarely contains the same speaker producing multiple accents. Additional supervision, such as explicit accent and speaker classification, may be required to further disentangle these factors.

\section{Discussion}

We observed a systematic delay of the optimal accessible layer relative to the recoverability peak. As discussed in \cite{yeh2024estimating}, earlier layers retain more complete acoustic information that can be effectively extracted by a strong HiFiGAN vocoder, whereas later layers gain phonetic discriminability from ASR-related objectives,

Across both metrics, accent information follows a distinct yet intermediate trend between phonetic and speaker information. This aligns with linguistic intuition: accent is partially realised through systematic phonetic variation and partially through speaker-specific pronunciation patterns.

Layer selection enables accent control precisely because accent, speaker, and phonetic information are distributed differently across layers. Consequently, a single "style layer", as suggested by the content–style token design in Vevo, is insufficient to jointly capture multiple speech attributes such as content, accent, and emotion. A more thorough investigation of more speech attributes, and how their information are represented across multiple layers and representations, is therefore necessary to support better controllable speech generation.

Finally, our findings provide potential explanations for the hallucinated accents by ZS-TTS systems \cite{zhong2025accentbox}. Many ZS-TTS systems rely on representations (often deep or supervised layers) where accent recoverability is already attenuated, encouraging the model to guess accent patterns in inference or default to mainstream accents in training data.

\section{Conclusion}

In this work, we propose a framework for evaluating discrete speech representation tokens (DSRTs) from both recoverability and accessibility perspectives, in terms of accent, speaker, and phonetic information encoded, with a focus on providing guidance for accent generation and control. Based on the results using this framework on HuBERT- and Whipser-based DSRTs, we propose content and content-accent tokens, achieving more controllable accent in voice conversion (VC) than existing works.

\clearpage

\section{Acknowledgments}

\ifcameraready
     This work was supported in part by the UKRI AI Centre for Doctoral Training in Responsible and Trustworthy in-the-world Natural Language Processing (Grant EP/Y030656/1), UKRI Centre for Doctoral Training in Natural Language Processing (Grant EP/S022481/1), School of Informatics, and School of Philosophy, Psychology \& Language Sciences, the University of Edinburgh. We would like to thank Dr. Hao Tang and Dr. Tianzi Wang for useful discussion.
\else
     [DEDACTED]
\fi

\section{Generative AI Use Disclosure}

During the preparation of this manuscript, the authors used ChatGPT (version 5.2 and 5.3) for language editing and readability improvement. The authors subsequently reviewed and revised the manuscript and take full responsibility for the final content.

\bibliographystyle{IEEEtran}
\bibliography{mybib}

\end{document}